\documentclass[12pt,a4paper]{article}
\newenvironment{keywords}{Keywords:}{}
\usepackage{fullpage}

\usepackage[utf8]{inputenc}
\usepackage{amsmath,amsfonts}
\usepackage{stmaryrd} 
\usepackage[usenames,dvipsnames,svgnames,table]{xcolor}
\usepackage{graphicx}
\usepackage{subfig}
\usepackage{sidecap}
\usepackage{natbib}
\usepackage{xfrac}
\usepackage{siunitx}
\usepackage{bm}

\pdfoutput=1
\usepackage{hyperref}
\hypersetup{
    colorlinks=true,
    linkcolor=black,
    citecolor=black,
    filecolor=black,
    urlcolor=black,
}
\usepackage[noabbrev]{cleveref}

\newcommand{\T}[1]{{\bm{{#1}}}}
\newcommand{\parm}{\mathord{\color{black!33}\bullet}}
\newcommand{\pdiff}[2]{\frac{\partial#1}{\partial#2}}

\newcommand{\psdiff}[2]{\sfrac{\partial#1}{\partial#2}}
\newcommand{\dif}{\,\mathrm{d}}

\newcommand{\jump}[1]{\left\llbracket\,#1\,\right\rrbracket}
\newcommand{\avg}[1]{\left\{\!\left\{ #1 \right\}\!\right\}}

\newcommand{\Hdiv}{H\textsuperscript{div}}
\newcommand{\norm}[1]{\left\lVert#1\right\rVert}

\newcommand{\vel}{\T{u}}
\newcommand{\wvel}{\T{w}}
\newcommand{\vtest}{\T{v}}
\newcommand{\pos}{\T{x}}
\newcommand{\grav}{\T{g}}

\newcommand{\domain}{\Omega}
\newcommand{\ddomain}{{\partial\Omega}}

\newcommand{\element}{K} 
\newcommand{\delement}{{\partial{}\element}}
\newcommand{\tess}{\mathcal{T}}
\newcommand{\skel}{\mathcal{S}}
\newcommand{\face}{F} 
\newcommand{\facet}{\face}

\newcommand{\normal}{\T{n}}

\begin{document}

\title{Slope limiting the velocity field in a discontinuous Galerkin divergence free two-phase flow solver}

\author{
 Tormod Landet%
   \thanks{Corresponding author. Email: tormodla@math.uio.no\vspace{6pt}}%
 , %
 Kent-Andre Mardal %
 and %
 Mikael Mortensen%
 \\%
 \vspace{6pt}%
 {\em{Matematisk Institutt, Moltke Moes vei 35, 0851 Oslo, Norway}}
}

\maketitle

\begin{abstract}
Solving the Navier-Stokes equations when the density field contains a large sharp discontinuity---such as a water/air free surface---is numerically challenging. Convective instabilities cause Gibbs oscillations which quickly destroy the solution.
We investigate the use of slope limiters for the velocity field to overcome this problem in a way that does not compromise on the mass conservation properties. The equations are discretised using the interior penalty discontinuous Galerkin finite element method that is divergence free to machine precision.

A slope limiter made specifically for exactly divergence free (solenoidal) fields is presented and used to illustrated the difficulties in obtaining convectively stable fields that are also exactly solenoidal. The lessons learned from this are applied in constructing a simpler method based on the use of an existing scalar slope limiter applied to each velocity component.

We show by numerical examples how both presented slope limiting methods are vastly superior to the naive non-limited method.
The methods can solve difficult two-phase problems with high density ratios and high Reynolds numbers---typical for marine and offshore water/air simulations---in a way that conserves mass and stops unbounded energy growth caused by the Gibbs phenomenon.


\end{abstract}

\section{Introduction}

The incompressible and variable density Navier-Stokes equations for the unknown velocity $\vel$ and pressure $p$, with gravity $\grav$ and spatially varying density $\rho$ and viscosity $\mu$,
\begin{align}
 \rho \left( \pdiff{\vel}{t} + (\vel\cdot\nabla) \vel \right) &= \nabla\cdot\mu\left(\nabla \vel + (\nabla\vel)^T\right) - \nabla p + \rho \grav, \label{eq:nsmom}\\
 \nabla\cdot \vel &= 0, \label{eq:nssol}\\
 \pdiff{\rho}{t} + \vel\cdot\nabla \rho &= 0, \label{eq:nsdens}
\end{align}
are used in applications where the fluid velocity is much smaller than the Mach number and where gravity effects are important such as for internal and surface gravity waves. Unfortunately, numerical problems will immediately occur when using \cref{eq:nsmom,eq:nssol,eq:nsdens} to study air/water free surface physics with a higher order solution method. These problems are due to the sharp factor 1000 jump in momentum across the liquid/gas interface which causes Gibbs oscillations in the velocity field, even though the true velocity has no discontinuity at the interface. The energy in the velocity field will eventually blow up to destroy the solution if this non-linear convective instability is not handled with care.

The convergence of numerical approximations to the Navier-Stokes equations with variable and potentially discontinuous density and viscosity was studied by \citet{liu_convergence_2007}. They show that if the continuous problem has an unique solution then a stable discontinuous Galerkin (DG) discretisation with a piecewise constant density field will converge to that solution. This work presents such a stable DG discretisation. Stable fractional step methods for solving the equations have been studied by \citet{guermond_projection_2000,pyo_gaugeuzawa_2007,guermond_splitting_2009}. Our results are computed using a direct solver on the coupled velocity and pressure system in order to eliminate any fractional step splitting errors from influencing the conclusions.


In this work we will use a piecewise constant density field and the volume of fluid, VOF, method by \citet{hirt_volume_1981} for evolving this density field in time while maintaining a sharp interface. In the VOF method a transport equation for the density is solved for a normalised so called \emph{colour function}, $C\in[0,1]$, which is linearly related to the density. Among the most commonly used variations are the algebraic VOF schemes CICSAM \citep{ubbink_numerical_1997} and HRIC \citep{muzaferija_two-fluid_1998}. The level set method \citep{osher_fronts_1988} is another alternative for tracking free surfaces. Modern versions exist that significantly improve on the historical mass conservation problems of the level set method---see, e.g., \citet{olsson_conservative_2005,Toure_2016_level_set_mass_cons}.

For most industrial applications the above methods are implemented in solvers based on the finite volume method where it is possible to ensure convective stability by appropriate use of flux limiters, see, e.g., the illustrative discussions and diagrams of \citet{sweby_high_1984} and \citet{leonard_simple_1988}. Flux limiters based on TVD or ENO/WENO schemes are applied in the solution of the momentum equation, and some sort of stable interface sharpening scheme such as CICSAM or HRIC is used for the flux of density. Finite volume methods can be partially extended to higher order by use of larger geometrical stencils, but this approach is not trivial to implement when using irregular grids, so in practice it is common that only immediate neighbours are used and most schemes are hence low order.

For irregular geometries there are two major research directions that look to enable higher order methods: immersed boundary methods and finite element methods. Immersed boundary methods use regular background grids and perform special treatment of grid cells near or inside objects embedded in the computational domain \citep{peskin_immersed_2002}. Finite element methods use basis functions with local support to enable high order approximating functions on irregular meshes. This work is based on a discontinuous Galerkin (DG) finite element method (FEM). DG-FEM has two main advantages over continuous Galerkin methods, the ease of using an upwind flux limiter for stabilising the linear convective instabilities and the option to create exactly divergence free and mass conserving numerical schemes \citep{cockburn_local_2004,cockburn_locally_2005}. There is one important drawback, and that is the significantly increased number of degrees of freedom.
A variation of the scheme from \citet{cockburn_locally_2005} is used, where instead of using a local discontinuous Galerkin, LDG, treatment of the elliptic term, a symmetric interior penalty, SIP, treatment \citep{arnold_interior_1982} is employed.

When using higher order approximating polynomials it is no longer sufficient to use only a flux limiter to obtain convective stability---a slope limiter must also be included in the method \citep{cockburn_rungekutta_1998,cockburn_runge-kutta_2001,kuzmin_vertex-based_2010}. The flux limiter ensures that the cell average values are bounded. When using higher order basis functions, the solution can go out of bounds in localized regions due to steep gradients inside each cell. A slope limiter prevents this by flattening steep slopes near discontinuities while leaving the solution untouched near smooth maxima so that the method's high order accuracy is retained. While a flux limiter is an implicit part of the equation system and has its stabilising effect included in the results from the linear equation solver, a slope limiter is typically applied to the resulting function as an explicit post-processing operator. Other options could be to use non-linear diffusion to combat the non-linear instability, or to apply spectral filtering, see, e.g., \cite{michoski_comparison_2016,zingan_implementation_2013}.


This paper starts with a description of the Gibbs instability and the discontinuous Galerkin method employed for solving the variable density Navier-Stokes equations in \cref{sec:method_gibbs,sec:method_prelims,sec:method_dg,sec:method_hdiv}. Slope limiting is then presented; first the hierarchical Taylor based slope limiter by \citet{kuzmin_vertex-based_2010,kuzmin_slope_2013} in \cref{sec:method_ht}, and then the possibility of constructing a vector field slope limiter that leaves the resulting velocity field both solenoidal and free from local maxima is explored in \cref{sec:method_sol}. After showing that it is likely not possible to obtain a single field that is both solenoidal and stable, separate limiting of the convecting and the convected velocity fields is introduced. Two alternatives are presented in \cref{sec:sol_lim_alg}; both ensure solenoidal convecting velocities while keeping convective stability. Readers familiar with DG methods and slope limiting can start at \cref{sec:method_sol}, though the presentation builds directly on the preceding methods, so some referring back may be needed. Results from numerical tests are shown in \cref{sec:results} and discussion and concluding remarks can be found in \cref{sec:discussion}.

\section{The numerical method}
\label{sec:method}

\subsection{Instabilities}
\label{sec:method_gibbs}

\begin{figure}[htb]
  \centering
  \subfloat[Initial condition]{%
    \label{fig:problem_a} %
    \includegraphics[width=0.3\textwidth]{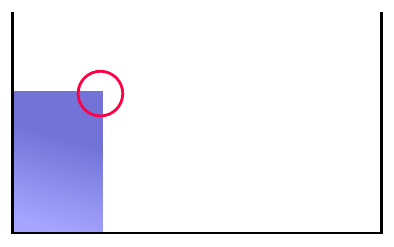}}
  ~
  \subfloat[Pinching]{%
    \label{fig:problem_b} %
    \includegraphics[width=0.3\textwidth]{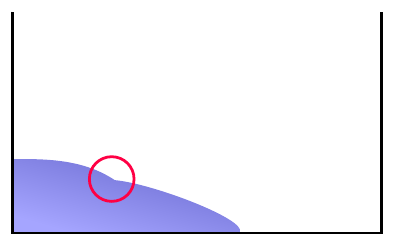}}
  ~ 
  \subfloat[Corner impact]{%
    \label{fig:problem_c} %
    \includegraphics[width=0.3\textwidth]{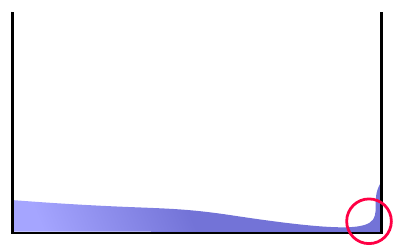}}
  \caption{Illustration of typical numerical problems}
  \label{fig:problems}
\end{figure}

The most common numerical instabilities related to handling of large density jumps are illustrated in \cref{fig:problem_a,fig:problem_b,fig:problem_c}. 
A `block' of water starts at rest in a box filled with air. Already in the very first time step (\cref{fig:problem_a}) the solution will start to blow up if one does not either apply smoothing to the density field or stabilise the numerical scheme. Smoothing the initial field is not sufficient as the interface may pinch and become non smooth (\cref{fig:problem_b}), so continuous smoothing of the density field is necessary if this approach is selected. A level set method is a natural way to implement a smoothed density field, see, e.g., \citet{unverdi_front-tracking_1992,sussman_level_1994}. In this work we will not apply stabilisation through smoothing. Our aim is to decouple the stability of the method from the treatment of the density field.

\Cref{fig:problem_c} illustrates a situation where the solenoidal properties of $\vel$ are important. A divergence free convecting velocity is required for the stability of the density transport \cref{eq:nsdens} and any divergence quickly becomes a problem in difficult situations such as the corner impact where in our experience either mass loss or unbounded densities will occur if care is not taken to ensure that the convecting velocity used for density transport is solenoidal.

\subsection{Preliminaries}
\label{sec:method_prelims}

The notation used in this paper is relatively standard. 
When looking at a facet between two cells (finite elements) we will denote one of the cells $K^+$ and the other $K^-$ in an arbitrary, but repeatable manner. Function values in each cell will be given the same superscripts to distinguish between the values on opposite sides of the facet in the discontinuous space. The average and jump operators across an internal facet are defined as 
\begin{align}
\avg{u} &= \frac{1}{2}(u^+ + u^-), \\
\jump{u} &= u^+ - u^-, \\
\jump{\T{u}}_\normal &= \T{u}^+\cdot\normal^+ + \T{u}^-\cdot\normal^-.
\end{align}

Vector terms such as the velocity $\vel$ and gravity $\grav$ will be denoted with a bold font while scalars such as the pressure $p$, the density $\rho$, and the dynamic viscosity $\mu$ are shown in italics. Projection operators are written in blackboard bold. An example is the projection $\mathbb{D}$ into a solenoidal vector space. Sets are written in a calligraphic typeface. Fluxes of a quantity, used in DG facet integrals, are marked with a circumflex accent over the quantity, e.g., $\hat{\vel}$, $\hat{p}$. An additional superscript is added when different definitions of the flux are used in different parts of the weak form, e.g., $\hat{\vel}^\wvel$ and $\hat{\vel}^p$, the flux of velocity related to convection and the flux of velocity related to the incompressibility constraint. 

Let $\partial K$ denote the boundary of element $K$. $\Gamma_I$ is the inlet portion of the domain boundary $\ddomain$ where $\vel\cdot\normal<0$ for an outwards pointing normal, $\normal$. The set of all grid cells is $\tess$ (the tessellation), and the set of all facets is $\skel$ (the mesh skeleton). The set of outside facets is $\skel_O=\skel\cap\ddomain$ and the set of inside facets shared between two cells is $\skel_I=\skel\setminus\skel_O$.

For a facet on the boundary, $\facet_b\in\skel_O$, let the boundary cell be denoted $K^+$ with regards to the facet and hence   $\normal^+\cdot\jump{\vtest}=\normal^+\cdot\vtest^+=\normal\cdot\vtest$ on $F_b$. We set terms related to $K^-$ to zero and also set the average to be the $K^+$ value $\avg{\vtest}=\vtest$ on $\skel_O$.

Let $P_k(\element)$ denote the space of polynomials of order $k$ on an element and $P_k(\facet)$ denote the space of polynomials of order $k$ on a facet. The basis functions in $P_k(\element)$ are discontinuous across facets and functions in $P_k(\facet)$  are discontinuous across edges (vertices in 2D). Let the superscript $\parm^{n+1}$ denote a value at time step  $t=\Delta t(n+1)$. For nabla the conventions $(\nabla\vel)_{ij}=\partial_j\vel_i$ and $(\nabla\cdot\T{\sigma})_i=\partial_j\sigma_{ij}$ are used. 

\subsection{Discretisation}
\label{sec:method_dg}

We will approximate the unknown functions in space by using discontinuous Lagrange polynomial function spaces with polynomial order $k=2$ for the velocity in $d=2$ spatial dimensions. Let the Galerkin test functions for $\{\vel, p, \rho\}$ be denoted respectively $\{\vtest, q, r\}$. The discontinuous function spaces are not restricted at the boundaries---all boundary conditions will be imposed weakly---so the trial and test functions share spaces,
\begin{align}
\vel, \vtest &\in [P_k(\element)]^d, \notag \\
p   , q      &\in P_{k-1}(\element), \\ \notag
\rho, r      &\in P_0(\element).
\end{align}

\subsubsection{Two-phase density transport}
The transport \cref{eq:nsdens} for the density is modified by the introduction of a solenoidal convecting velocity field $\wvel$ which is close to $\vel$, but may not be identical as will be explained in \cref{sec:method_hdiv,sec:sol_lim_alg}. Boundary conditions are needed only on the inlet. The strong form can then be written
\begin{align}
\pdiff{\rho}{t} + \wvel\cdot\nabla \rho = 0 \qquad &\text{in}\ \domain, 
\label{eq:dens_strong_w} \\ \notag
\rho = \rho_I \qquad &\text{on}\ \Gamma_I.
\end{align}
In VOF methods the fluid properties are expressed in terms of an indicator function, $C\in[0,1]$. The true density and viscosity fields can easily be recovered from $C$ when the density and kinematic viscosity properties of the two fluids are known,
\begin{align}
\rho &= C \rho_\text{water} + (1 - C)\rho_\text{air}, 
\label{eq:rho_from_c}
\\
\mu &= \left[C \nu_\text{water} + (1 - C)\nu_\text{air}\right] \rho.
\label{eq:mu_from_c}
\end{align}
In order to compute $C$ the transport \cref{eq:dens_strong_w} for the density is modified by inserting \cref{eq:rho_from_c}. Now the solution $C^{n+1} \in P_0(\element)$ can be found by expressing the resulting equation on weak form and integrating by parts,
\begin{align}
&\int_\tess\frac{1}{\Delta t}(\gamma_1 C^{n+1} + \gamma_2 C^{n} + \gamma_3 C^{n-1})r\dif\pos
\label{eq:colour_weak} \\ \notag
&\qquad- \int_\tess C^{n+1}\wvel\cdot\nabla r\dif\pos
+ \int_\skel\hat C^{n+1}\,\wvel\cdot\normal^+\jump{r}\dif s = 0,
\end{align}
where we have assumed that the convecting velocity $\wvel$ is \Hdiv-conforming such that the flux is continuous across facets, $\jump{\T{w}}_\normal=0$.

A second order backwards differencing formulation, BDF2, is used for time integration. The parameters are $\{\gamma_1, \gamma_2, \gamma_3\} = \{\sfrac{3}{2}, -2, \sfrac{1}{2}\}$. The BDF2 method is monotonicity preserving when started using a backward Euler step \citep{hundsdorfer_2003}, though the time step required to preserve monotonicity is half that of backward Euler. The benefit is that the method is second order---and small time steps are anyhow required to keep the interface sharp \citep{muzaferija_two-fluid_1998}. Second order extrapolation is used for the convective velocity,
\begin{equation}
\wvel = 2\wvel^{n} - \wvel^{n-1},
\label{eq:explicit_w}
\end{equation}
which makes $\rho^{n+1}$ independent of the computed velocity at time $t=(n+1)\Delta t$. The density transport equation is hence uncoupled from the momentum equation.

For the density flux, $\hat C^{n+1}$, the most stable choice is to take the upwind value, which means that the boundary condition $\hat\rho=\rho_I$ is used to determine $\hat C^{n+1}$ on inlet facets, $\skel \cap \Gamma_I$.
For the internal facets the term related to the flux can be written on upwind form as 
\begin{align}
\hat C^{n+1}\,\wvel\cdot\normal^+ = \jump{C^{n+1}\frac{1}{2}(\wvel\cdot\normal + |\wvel\cdot\normal|)},
\label{eq:rho_flux_upwind}
\end{align}
and by replacing the plus by a minus in \cref{eq:rho_flux_upwind} the downwind flux can similarly be computed. Using the upwind and downwind fluxes on each facet a downwind-blended compressive interface flux can be applied to the density transport such as CICSAM \citep{ubbink_numerical_1997} or HRIC \citep{muzaferija_two-fluid_1998}. Such blended fluxes sharpen the interface between the two fluid layers, but remain convectively stable unlike downwind or central fluxes. Both CICSAM and HRIC are algebraic VOF methods which define facet-wise blending factors to combine the upwind and downwind fluxes into one linearly stable flux. The results in \cref{sec:results} are computed using the HRIC method. 

One important note is that standard VOF flux limiters from finite volume methods ensure that the $C$ field remains sharp and bounded based on a convecting velocity field that is piecewise constant on each facet. Such a field can easily be computed from $\wvel$ and as long as $\wvel$ is solenoidal, so is the piecewise constant field. It is this field that is used when solving for $C^{n+1}$ since $\wvel$ is only needed on the facets---for $r\in P_0(\element)$ the volume integral term containing $\wvel$ in \cref{eq:colour_weak} is identically zero.

\subsubsection{The variable density Navier-Stokes equations}
The strong form of the variable density Navier-Stokes equations, where the convecting velocity is replaced by $\wvel$, can be written
\begin{align}
 \rho \left( \pdiff{\vel}{t} + (\wvel\cdot\nabla) \vel \right) &= \nabla\cdot\mu(\nabla \vel + \left(\nabla\vel)^T\right) - \nabla p + \rho \grav \qquad &\text{in}\ \domain,
 \label{eq:ns_with_w_conv}
 \\ \notag
 \nabla\cdot \vel &= 0 \qquad &\text{in}\ \domain, 
 \\ \notag
 \vel &= \vel_D \qquad &\text{on}\ \Gamma_D,
 \\ \notag
  \pdiff{\vel}{n} &= \T{a} \qquad &\text{on}\ \Gamma_N,
\end{align}
where $\Gamma_D$ and $\Gamma_N$ are the parts of the boundary where Dirichlet and Neumann boundary conditions are applied respectively; $\Gamma_N=\ddomain\setminus\Gamma_D$. 
Dirichlet boundary conditions can be enforced on external facets due to the elliptic viscosity term, $\nabla\cdot\mu(\nabla \vel + \left(\nabla\vel)^T\right)$. The Navier-Stokes equations will be written on weak form, see \cref{eq:wf_ns_coupled1}, but first we will briefly discuss the symmetric interior penalty (SIP) method. First, the boundary condition, $\vel=\vel_D$, is written on weak form on an external facet $\facet$,
\begin{align}
\int_\facet \vel\cdot\vtest\dif s =  \int_\facet \vel_D\cdot\vtest\dif s, \label{eq:weakbc_vel}
\end{align}
and then---using the stabilisation scheme proposed by \citet{nitsche_uber_1971}---the test function $\vtest$ is replaced by a Petrov-Galerkin test function $\tilde\vtest=\kappa_\mu\vtest - \mu\left(\nabla\vtest + (\nabla\vtest)^T\right)\cdot\normal$. This method is extended to enforce continuity across internal facets which is necessary for stability \citep{arnold_interior_1982}. This gives 
\begin{align}
\int_\facet\kappa_\mu\jump{\vel}\cdot\jump{\vtest}\dif s
- \int_\facet (\avg{\mu\left(\nabla\vtest + (\nabla\vtest)^T\right)}\cdot\normal^+)\cdot\jump{\vel}\dif s
 = 0.
 \label{eq:weakbc_vel_internal}
\end{align}
where $\kappa_\mu$ is a penalty parameter which must be sufficiently large to ensure stability. The analyses in \citet{epshteyn_estimation_2007} and \citet{shahbazi_high-order_2007} guide us in defining the penalty parameter as a function of the minimum and maximum dynamic viscosities, $\mu_\mathrm{min}$ and $\mu_\mathrm{max}$, the order $k$ of the approximating polynomials, and the surface area $S_K$ and volume $V_K$ of each cell $K$,
\begin{align}
\kappa_\mu = 3\,\frac{\mu_\mathrm{max}^2}{\mu_\mathrm{min}}\,k(k+1)\,\max_K\left(\frac{S_K}{V_K}\right).
\label{eq:penalty_estimate}
\end{align}

The same scheme is used for the left-hand side of the momentum equation as for the density transport equation, but a pure upwind flux is used without any downwind blending for the convective term. To avoid overloading the notation we now drop the $\parm^{n+1}$ superscript on the unknown quantities. The upwind flux related to convection can then be written
\begin{align}
\hat\vel^\wvel\,\wvel\cdot\normal^+ = \jump{\vel\frac{1}{2}(\wvel\cdot\normal + |\wvel\cdot\normal|)}.
\label{eq:vel_flux_upwind}
\end{align}

Both the pressure gradient and the viscosity on the right-hand side of the momentum equation are integrated by parts. The resulting weak form is a direct combination of the LDG Navier-Stokes method by \citet{cockburn_locally_2005} and the SIP diffusion method by \citet{arnold_interior_1982}. Both these references contain more details and proofs of stability. The stability of the convective and the diffusive terms are not interconnected, so replacing the LDG elliptic operator with the SIP version is unproblematic. The treatment of the momentum transport and the inter-cell continuity, both described above, are easily recognisable in the complete weak form,
{\allowdisplaybreaks
\begin{flalign}
&\int_\tess\frac{\rho}{\Delta t}(\gamma_1\vel + \gamma_2\vel^{n} + \gamma_3\vel^{n-1})\vtest\dif\pos
\label{eq:wf_ns_coupled1}
\\ \notag
&\qquad-\ \int_\tess \vel \cdot\nabla\cdot (\rho\vtest\otimes\wvel)\dif\pos
 \ +\ \int_\skel \wvel\cdot\normal^+\,\hat\vel^\wvel\cdot\jump{\rho\vtest}\dif s
\\ \notag
&\qquad+\ \int_\tess \mu\left(\nabla\vel + (\nabla\vel)^T\right):\nabla\vtest\dif\pos
\ +\ \int_{\skel_I}\kappa_\mu\jump{\vel}\cdot\jump{\vtest}\dif s
\\ \notag
&\qquad-\ \int_\skel (\avg{\mu\left(\nabla\vel + (\nabla\vel)^T\right)}\cdot\normal^+)\cdot\jump{\vtest}\dif s
\\ \notag
&\qquad-\ \int_{\skel_I} (\avg{\mu\left(\nabla\vtest + (\nabla\vtest)^T\right)}\cdot\normal^+)\cdot\jump{\vel}\dif s
\\ \notag
&\qquad-\ \int_\tess p\, \nabla\cdot\vtest\dif\pos
 \ +\ \int_\skel\hat{p}\,\normal^+\cdot\jump{\vtest}\dif s
\ =\ \int_\tess\rho\,\grav\dif\pos.
\end{flalign}}
where the flux of pressure is taken as $\hat{p}=\avg{p}$. 

The continuity \cref{eq:nssol} is also integrated by parts using $\hat{\vel}^p=\avg{\vel}$ as the flux related to the incompressibility constraint,
\begin{align}
&\int_\skel \hat{\vel}^p\cdot\normal^+\jump{q} \dif s
 \ -\ \int_\tess \vel\cdot\nabla q\dif\pos
 \ =\ 0.
\label{eq:wf_ns_coupled2}
\end{align}

\paragraph{Dirichlet boundaries}
On the inflow part of the Dirichlet boundary take $\hat{\vel}^\wvel=\vel_D$, and on the outflow part take $\hat{\vel}^\wvel=\vel$, i.e., the upwind values are used. On the whole Dirichlet boundary let $\hat{\vel}^p=\vel_D$ and $\hat{p}=p$. The viscous penalty and symmetrisation terms from \cref{eq:weakbc_vel_internal} can be used also on the domain boundary by replacing $\vel^-$ by $\vel_D$. As noted in the references used to define $\kappa_\mu$ above, the best choice is to use twice the amount of penalisation on external facets compared to the interior facets. This leads to external boundary integrals
\begin{align}
\int_{\Gamma_D} 2 \kappa_\mu (\vel - \vel_D)\cdot\vtest \dif s
-\int_{\Gamma_D} (\mu\nabla\vtest\cdot\normal)\cdot(\vel-\vel_D)\dif s
= 0. \label{eq:weakbc_vel_external}
\end{align}

\paragraph{Neumann boundaries}
For the Neumann boundaries we take $\hat{\vel}^\wvel=\vel$, $\hat{\vel}^p=\vel$ and $\hat{p}=p$. The extra viscous terms for symmetrisation and penalty are removed and only the normal integration by parts terms are left. The surface term becomes
\begin{align}
-\int_{\Gamma_N} \mu\,\T{a}\cdot\vtest\dif s.
\end{align}

Pure Neumann boundary conditions will not be used, but we will use Dirichlet for one velocity component and Neumann for the other to implement free-slip boundary conditions on planes that are parallel to the axes. The definitions above can easily be split into component-wise treatment of boundary conditions.

\paragraph{Solution algorithm} \ \\

The Navier-Stokes equations and the density transport equation are solved in a decoupled manner for each time step in the following order:
\begin{enumerate}
 \item Compute an explicit convecting velocity $\wvel^{n+1}$ by use of \cref{eq:explicit_w}.
 \item Find $C^{n+1} \in P_0(\element)$ such that \cref{eq:colour_weak} is satisfied.
 \item Compute the density $\rho^{n+1}$ and viscosity $\mu^{n+1}$ by use of \cref{eq:rho_from_c,eq:mu_from_c}.
 \item Use the computed coefficient fields $\wvel^{n+1}$, $\mu^{n+1}$ and $\rho^{n+1}$ to find $\vel^{n+1} \in [P_k(\element)]^d$ and $p^{n+1} \in P_{k-1}(\element)$ from \cref{eq:wf_ns_coupled1,eq:wf_ns_coupled2}.
\end{enumerate}

\subsection{\Hdiv{} projection of the velocity field}
\label{sec:method_hdiv}

In finite element methods for solving the Navier-Stokes equations it is common to impose the incompressibility criterion, $\nabla\cdot\vel=0$, weakly by multiplying with a scalar test function, $q$, and integrating over the domain. This term, $\int_\domain \nabla\cdot\vel\,q\dif \pos$, will appear directly in a coupled solver and as the right-hand side in the Poisson equation for the pressure in a pressure correction fractional step scheme such as the commonly used incremental pressure correction scheme.

Imposing the incompressibility criterion weakly in the space of the pressure is sufficient for stability, but one must require exact incompressibility to locally conserve mass and momentum. Below we explain the core ideas of the method presented in \citet{cockburn_locally_2005}. By using this method the resulting divergence will be zero almost to machine precision, approximately $10^{-13}$ on each cell in our tests. We calculate the cell-wise error by computing the integrated absolute value of the divergence internally in each cell, $\int_\element |\nabla\cdot\vel| \dif \pos$, and add to it the error in flux continuity between cells on each connected facet, $\int_\delement |\jump{\vel\cdot \normal}| \dif s$.

Let us start by noting that \Hdiv{}-conforming finite elements exist, and that such elements of a given polynomial order will be subspaces of the fully discontinuous elements of the same order with the same cell geometry. Such elements impose continuity of normal fluxes across facets and hence have fewer global degrees of freedom. We follow \citet{cockburn_locally_2005} and define a projection from our fully discontinuous velocity $\vel$ into a velocity $\wvel =\mathbb{P}\vel$ that exist in a space of polynomials that are consistent with the \Hdiv{}-conforming elements.


The projection operator $\wvel=\mathbb{P}\vel$ is defined in a cell-wise manner. This local projection is hence very fast and consists of finding $\wvel \in [P_k(\element)]^d$ such that
\begin{align}
\int_\facet \wvel \cdot\normal \, v_1 \dif s &= \int_\facet \hat{\vel}^p \cdot \normal \, v_1 \dif s 
& \forall v_1 \in P_k(\facet), \facet \in \delement,
\label{eq:bdm_proj1}
\\
\int_\element \wvel \cdot \T{v}_2 \dif x &= \int_\element \vel \cdot \T{v}_2 \dif x
& \forall \T{v}_2 \in \T{N}_{k-1}(\element).
\label{eq:bdm_proj2}
\end{align}
The first equation ensures the continuity of the normal velocities across each facet. The continuity stems from using a single valued flux which is here $\hat{\vel}^p=\avg{\vel}$ on internal facets and $\hat{\vel}^p=\vel_D$ on external facets ($\hat{\vel}^p=\vel$ on $\Gamma_N$). This flux is consistent for continuous velocity fields.

In \cref{eq:bdm_proj2} the space $\T{N}_{k-1}(\element)$ is the Nédélec $H(\text{curl})$ element of the first kind of order $k-1$, see, e.g., \citet{kirby_common_unusual_2012,nedelec_new_mixed_family_1986}. The dimension of the Brezzi-Douglas-Marini (BDM) element $P_k(\facet)\times\T{N}_{k-1}(\element)$ is the same as that of the Discontinuous Lagrange DG$k$ element on each cell \citep{brezzi_mixed_1991}. It is hence possible to form square projection matrices between the two spaces in each cell. The projection $\mathbb P \vel$ is not square globally since the degrees of freedom related to $v_1$ are shared between exactly two cells on all internal facets. This leads to the test space having fewer global degrees of freedom than the fully discontinuous trial space.

The properties of this BDM-like projection used as a velocity post-processing step in a discontinuous Galerkin method is given by \citet{cockburn_locally_2005}. The projection gives both a continuous flux $\wvel\cdot\normal$ on all inter-element facets and also ensures that the total flux across each individual cell's facets is zero. Fulfilling these two criteria is what we mean by exact incompressibility.

\section{Slope limiting}
\label{sec:slopelim}

The numerical diffusion due to upwinding in the spatial DG scheme is sufficient to stabilise convective operators and avoid spurious oscillations when piecewise constant approximating functions are employed. One way to stabilise convective operators when using higher order approximating functions is to employ slope limiters \citep{cockburn_rungekutta_1998}.

\subsection{The hierarchical Taylor polynomial based slope limiter}
\label{sec:method_ht}

The hierarchical vertex based slope limiter by \citet{kuzmin_vertex-based_2010,kuzmin_slope_2013} is used to remove high frequency oscillations near discontinuities when solving an equation containing a convective term for a scalar quantity $\phi$. This slope limiter is based on using discontinuous Taylor function spaces. We are using Lagrange polynomials, so the first step is to project the unknown function $\phi$ from the discontinuous Lagrange function space to a function $\phi_t$ in the discontinuous Taylor function space. This projection $\phi_t=\mathbb{T}\phi$ is local to each cell and can be applied and inverted exactly by a single matrix vector product in each cell, so the cost of converting back and forth is negligible.

The discontinuous Taylor function space is slightly altered from the standard definition by use of cell averages. The expansion
\begin{align}
\label{eq:taylorbasis}
\phi_t(x,y) = & \bar{\phi}
         + \left.\pdiff{\phi}{x}\right|_c \alpha_1 (x - x_c)
         + \left.\pdiff{\phi}{y}\right|_c \alpha_1 (y - y_c) +
         \\ \notag
         &\left.\pdiff{^2\phi}{x^2}\right|_c \alpha_2 \left[ \frac{(x-x_c)^2}{2} - \overline{\frac{(x-x_c)^2}{2}}\right]
         + \left.\pdiff{^2\phi}{y^2}\right|_c \alpha_2 \left[ \frac{(y-y_c)^2}{2} - \overline{\frac{(y-y_c)^2}{2}}\right] +
         \\ \notag
         &\left.\pdiff{^2\phi}{x\partial y}\right|_c \alpha_2 \left[ \frac{(x-x_c)(y-y_c)}{2} - \overline{\frac{(x-x_c)(y-y_c}{2}}\right]
\end{align}
is used to describe a second order polynomial function on a triangle
where the six coefficients are $\{\bar{\phi}, \left.\psdiff{\phi}{x}\right|_c, \cdots, \left.\psdiff{^2\phi}{x\partial y}\right|_c\}$. The number of coefficients is the same as the number of nodes in a second order Lagrangian function space on a triangle which ensures that the transformation matrix resulting from $\mathbb{T}$ is square on each cell. The same is also true for tetrahedra and for higher and  lower polynomial orders.

Over-lined terms such as $\bar{\phi}$ and $\overline{\sfrac{(x-x_c)^2}{2}}$ denote cell averages and the restriction $\left.\cdot\right|_c$ signifies evaluation in the cell centre. Slope limiters $\alpha_i\in[0,1]$ will be applied to the $i$'th derivative terms as shown in \cref{eq:taylorbasis} using the same $\alpha$-factors for all derivatives of the same order. How to compute $\alpha_i$ is described below. As long as no slope limiting is performed ($\alpha_i=1$), we have $\phi=\mathbb{T}^{-1}\mathbb{T}\phi$.

To find $\alpha_i$ we follow \citet{kuzmin_slope_2013} and define three bi-linear functions on each cell,
\begin{align}
\tilde{\phi}_0(x,y) &= \bar{\phi} + \tilde{\alpha}_1\left[\left.\pdiff{\phi}{x}\right|_c(x-x_c) + \left.\pdiff{\phi}{y}\right|_c(y-y_c)\right],
\label{eq:htlimapprox1}
\\
\tilde{\phi}_x(x,y) &= \left.\pdiff{\phi}{x}\right|_c + \tilde{\alpha}_{2x}\left[\left.\pdiff{^2\phi}{^2x}\right|_c(x-x_c) + \left.\pdiff{^2\phi}{x\partial y}\right|_c(y-y_c)\right],
\label{eq:htlimapprox2}
\\
\tilde{\phi}_y(x,y) &= \left.\pdiff{\phi}{y}\right|_c + \tilde{\alpha}_{2y}\left[\left.\pdiff{^2\phi}{^2 y}\right|_c(y-y_c) + \left.\pdiff{^2\phi}{x\partial y}\right|_c(x-x_c)\right].
\label{eq:htlimapprox3}
\end{align}
The parameters $\tilde{\alpha}_i$ are intermediate slope limiters for the approximate function $\tilde{\phi}_i$. \Cref{eq:htlimapprox1,eq:htlimapprox2,eq:htlimapprox3} are used to linearly approximate $\phi$ and its first derivatives at each of the three vertices of the triangular cells.  These approximate vertex values should not form local maxima or minima when compared to the cell centre values of the linear representations in the surrounding cells.

Pick any one of the above linear approximations in \cref{eq:htlimapprox1,eq:htlimapprox2,eq:htlimapprox3} and call it $\tilde{\phi}(x,y)$. The corresponding value in the cell centre is denoted $\tilde{\phi}_c=\tilde{\phi}(x_c,y_c)$. Iterate through all cells in the mesh and for each cell consider each vertex and record the extremal $\tilde{\phi}_c$ in the cells that share the vertex. This gives allowable bounds $\tilde{\phi}_i^\mathrm{min}$ and $\tilde{\phi}_i^\mathrm{max}$ at vertex $i$ for the selected linear approximation in the given cell.
For each vertex $i$ located at $(x_i, y_i)$ one must ensure that $\tilde{\phi}_i = \tilde{\phi}(x_i,y_i)$ is bounded by the surrounding cell values, $\tilde{\phi}_i^\mathrm{min} \le \tilde{\phi}_i \le \tilde{\phi}_i^\mathrm{max}$. To find the maximum admissible value of $\tilde{\alpha}$ that ensures this, first for each vertex $i$ set
\begin{align}
\tilde{\alpha}_i = \begin{cases}
    \min\{1, \frac{\tilde{\phi}_i^\mathrm{max} - \tilde{\phi}_c}{\tilde{\phi_i}-\tilde{\phi}_c}\}       & \quad \text{if } \tilde{\phi_i}-\tilde{\phi}_c > 0,
    \\
 1       & \quad \text{if } \tilde{\phi_i}-\tilde{\phi}_c = 0,
 \\
  \min\{1, \frac{\tilde{\phi}_i^\mathrm{min} - \tilde{\phi}_c}{\tilde{\phi_i}-\tilde{\phi}_c}\}       & \quad \text{if } \tilde{\phi_i}-\tilde{\phi}_c < 0,\\
  \end{cases}
\end{align}
and then take the minimum of the calculated $\tilde{\alpha}_i$-values in each vertex to find $\tilde{\alpha}_1$, $\tilde{\alpha}_{2x}$ and $\tilde{\alpha}_{2y}$ for each cell by performing the above calculations for each of the three linear approximations in \cref{eq:htlimapprox1,eq:htlimapprox2,eq:htlimapprox3}. The last step is to calculate the final slope limiter coefficients for the second order derivatives,
\begin{align}
\alpha_2 = \min\{\tilde{\alpha}_{2x}, \tilde{\alpha}_{2y}\},
\end{align}
and, since one can expect higher regularity of the first order derivatives than the second order derivatives, take
\begin{align}
\alpha_1 = \max\{\tilde{\alpha}_1, \alpha_2\},
\end{align}
which at a smooth extremal point will stop any limiting from happening since $\alpha_2=1$ here, even if it is likely that $\tilde{\alpha}_1 < 1$.

Having found the slope limiter coefficients $\alpha_1$ and $\alpha_2$ for a given cell one can go back to the discontinuous Lagrange function space with the help of the projection $\mathbb{T}^{-1}$. If there were no spurious oscillations in the cell then the slope limiters coefficients should end up as $\alpha_i=1$ and the slope limiter projection---which we call $\mathbb{S}$---will hence be an identity transform, and we get $\phi^\mathrm{lim}=\mathbb{T}^{-1}\,\mathbb{S}\,\mathbb{T}\,\phi=\mathbb{T}^{-1}\,\mathbb{T}\,\phi=\phi$. This ensures that the order of spatial convergence is kept the same as the underlying DG scheme.

\subsection{On slope limiting of solenoidal fields}
\label{sec:method_sol}

Slope limiting of solenoidal vector fields is more complex than limiting scalar fields due to an increased number of invariants. For scalar fields the only invariant is that the average value in each cell must be unchanged after limiting. For vector fields the (i) divergence inside each element and the (ii) flux between neighbouring elements are new invariants in addition to the (iii) average value of each of the velocity component in each cell. Breaking these invariants means that the result is not solenoidal (i), the slope limiting post processor is not a non-local operation (ii) or that momentum is not preserved (iii).

In the following sections we will describe how these invariants can be used to reduce the number of unknowns in the cell-wise limiting problem from 12 to only 4 for a DG2 vector field, $\vel \in [P_2(\element)]^2$, on a triangle. We will then describe an optimisation method which can be used to decide the value of the remaining four unknowns in a way that minimises the tendency to produce local extrema. Removing all local extrema would render the method stable. We do not claim that the below method is optimal, but it does give strong indications that the four remaining unknowns in the cell-wise limiting problem are not sufficient to fully control the local extrema. Creating a single velocity field that is both solenoidal and stable is hence likely not possible by use of a cell-wise slope limiting process.

\subsubsection{A reduced basis for the solenoidal slope limiting problem}

Let the cell volume be denoted $V_\element$ and the area of facet $i$ be denoted $L_i$. Preserving the facet average of the flux between neighbouring cells gives a result which is \Hdiv-compatible in a DG0 sense, which is what is needed for mass conservation. The cell-wise invariants we require to be left unchanged by the slope limiting procedure are then
\begin{align}
\begin{aligned}[t]
R_i &= \frac{1}{L_i} \int_{\facet_i} \vel\cdot\normal\dif s
\ \ \forall i\in\{1, 2, 3\},
\end{aligned}
\qquad
\begin{aligned}[t]
U_j &= \frac{1}{V_\element} \int_\element \vel_j \dif x
\ \ \forall j\in\{1, 2\}.
\end{aligned}
\label{eq:invariants}
\end{align}

When applying this to a DG2 vector field on a triangle there are five invariants, three average fluxes, $R_i$, and two component averages, $U_j$. By not changing the average flux on each facet the total flux for each cell will still sum to zero, and it is still possible to have a solenoidal description of the vector field. 

The 12 degrees of freedom in a DG2 vector field on a triangle can be reduced. The divergence free solution will never span the full Lagrange space, but be restricted to the solenoidal subspace which has only 9 degrees of freedom, see e.g. \citet{baker_piecewise_1990}. This space is spanned by
\begin{align}
\binom{1}{0},\,   \binom{0}{1};     \quad\ 
\binom{y}{0},\,   \binom{0}{x},\,   \binom{x}{-y};       \quad\ 
\binom{y^2}{0},\, \binom{0}{x^2},\, \binom{x^2}{-2xy},\, \binom{-2xy}{y^2};
\label{eq:solpols}
\end{align}
which is here shown grouped into the two zeroth order, three first order and four second order vector valued polynomials.

A projection from the Lagrangian description of $\vel$ to the solenoidal description $\vel_s$ can be implemented as $\vel_s=\mathbb{D}\vel$ on a cell by cell level. To do this, first define local $x$ and $y$ coordinates with the origin in the centre of each cell and then form the cell-wise rectangular inverse projection $\vel=\mathbb{D}^{-1}\vel_s$ by evaluating the solenoidal polynomials from \cref{eq:solpols} in the location of the Lagrangian nodes. The least squares pseudo-inverse of this matrix is then $\mathbb{D}$. It is important to note that in our simulations the projection is lossless such that $\vel=\mathbb{D}^{-1}\mathbb{D}\vel$ since the \Hdiv{} projection along with the DG weak form of the momentum and continuity equations causes $\vel$ to be fully described by the solenoidal subspace. In general, we have $\mathbb{D}^{-1}\mathbb{D}\neq\mathbb{I}$, so this identity property does not hold for a generic vector field.

The velocity vector field can now be described by nine degrees of freedom in each cell, coefficients $s_1\cdots s_9$. Each of the nine coefficients is multiplied by the corresponding vector valued polynomials in \cref{eq:solpols}, numbered from left to right. The sum of these products form the complete vector valued polynomial field. 

We will further restrict the number of degrees of freedom by making use of the five invariants from \cref{eq:invariants}. Let the coefficients for the constant and linear polynomials be determined by the five invariants. The four quadratic terms are now the only degrees of freedom. Picking four arbitrary numbers, $\sigma_1$ to $\sigma_4$, for the coefficients $s_6$ to $s_9$, one can find corresponding constant and linear weights that make the final vector field keep the selected invariants. This is done by forming a 9x9 matrix system,
\begin{align}
\begin{bmatrix}
  q_{11}& q_{12}& q_{13}& q_{14}& q_{15}& q_{16}& q_{17}& q_{18}& q_{19} \\
  q_{21}& q_{22}& q_{23}& q_{24}& q_{25}& q_{26}& q_{27}& q_{28}& q_{29} \\
  q_{31}& q_{32}& q_{33}& q_{34}& q_{35}& q_{36}& q_{37}& q_{38}& q_{39} \\
  q_{41}& q_{42}& q_{43}& q_{44}& q_{45}& q_{46}& q_{47}& q_{48}& q_{49} \\
  q_{51}& q_{52}& q_{53}& q_{54}& q_{55}& q_{56}& q_{57}& q_{58}& q_{59} \\
  0& 0& 0& 0& 0& 1& 0& 0& 0 \\
  0& 0& 0& 0& 0& 0& 1& 0& 0 \\
  0& 0& 0& 0& 0& 0& 0& 1& 0 \\
  0& 0& 0& 0& 0& 0& 0& 0& 1 \\
\end{bmatrix}
\cdot 
\begin{bmatrix}
  \,s_1\, \\ s_2 \\ s_3 \\  s_4 \\ s_5 \\ s_6 \\ s_7 \\ s_8 \\ s_9
\end{bmatrix}
=
\begin{bmatrix}
  \, U_1 \, \\ U_2 \\ R_1 \\  R_2 \\ R_3 \\ \sigma_1 \\ \sigma_2 \\ \sigma_3 \\ \sigma_4
\end{bmatrix},
\label{eq:sigma_to_s}
\end{align}
where the first five rows contain appropriate weights $q_{ki}$ such that the correct integral from \cref{eq:invariants} is computed when the row is multiplied by the vector of coefficients, $s_i$. These weights can easily be computed by quadrature. By solving \cref{eq:sigma_to_s} the full set of nine solenoidal coefficients, $s_i$, can be recovered from the reduced set of four degrees of freedom, $\sigma_j$. The original 12 degrees of freedom of the Lagrangian velocity field $\vel$ can be recovered by use of the inverse projection $\vel=\mathbb{D}^{-1}\vel_s$.

We should note that we have no proof that the matrix in \cref{eq:sigma_to_s} is invertible for all possible cells, but in our extensive testing we have never found a cell where this was not the case. For other possible matrices where $\sigma_j$ do not control the quadratic terms in the solenoidal basis---but instead let's say that $\sigma_j$ are identical to the first four $s_i$---then it is easy to find cells where the condition number of the resulting matrix is very high, so the choice of the limited degrees of freedom $\sigma_j$ is important.


\subsubsection{Optimisation and cost functions}
\label{sec:sol_opt_cost}

We have used optimisation to determine the remaining four degrees of freedom by minimising a cost function in $\mathbb R^4$. This optimisation is performed for each cell to find the coefficients resulting in the lowest cost, which should correspond to the most stable solution with an appropriate choice of cost function. For each choice of cost function we get a new method with unique properties.


An ideal cost function will keep the limited solution very close to the original non-limited solution while still damping non-physical oscillations. To achieve this we have selected to first calculate a component-wise slope limited solution $\vel_\text{HT}$ by applying the hierarchical Taylor slope limiter described in \cref{sec:method_ht} to each of the velocity components. This makes the result frame dependent, which is not desirable, but sadly unavoidable when using a scalar field slope limiter for a vector field.

The $\vel_\text{HT}$ field is used as the target in the optimisation since it maintains the correct convergence order. This target field does not keep all the solenoidal invariants, so it is always slightly out of reach. The minimum and maximum allowable velocity component values are computed for a set of points along each cell boundary by considering the average values in the neighbouring cells just like in the hierarchical Taylor slope limiter. The range of allowable values is extended to include the already limited field $\vel_\text{HT}$ in order to avoid limiting at smooth maxima.

The cost function is implemented in two steps. First take the coefficients $s_i$ corresponding to the non-limited velocity field and use these in \cref{eq:sigma_to_s} to calculate the five invariants from \cref{eq:invariants}. Then, for a set of points $P_i\in\delement$ along the perimeter of a cell $\element$, compute for each velocity component $u_j\ (j\in[1,2])$ the current value $u_{ij}=u_j(P_i)$, the allowable minimum $u_{ij,\text{min}}$, the preferred target $u_{ij,\text{HT}}$, and the allowable maximum $u_{ij,\text{max}}$. The cost at point $P_i$ for velocity component $u_j$ is then the squared distance from the target normalised by the range of allowed values,
\begin{align}
C_{ij} = \left(\frac{u_{ij} - u_{ij,\text{HT}}}{u_{ij,\text{max}} - u_{ij,\text{min}}}\right)^2.
\end{align}
As the second step, if the current value $u_{ij}$ is higher than $u_{ij,\text{max}}$, add
\begin{align}
C_{ij,\kappa} = \kappa_L\left(\frac{u_{ij} - u_{ij,\text{max}}}{u_{ij,\text{max}} - u_{ij,\text{min}}}\right)^2 + \kappa_C,
\end{align}
and similar if $u_{ij} < u_{ij,\text{min}}$. If the current value is inside the stable bounds, then $C_{ij,\kappa} =  0$. We have used $\kappa_L=1.0$ and $\kappa_C= 1.0$ as additional penalties for going outside the stable zone. We have used the six Lagrangian nodes as the points $P_i$, i.e., the vertices and facet midpoints. The total cost for the cell $\element$ to be optimised is then calculated as
\begin{align}
C_\element = \sum_{i=1}^6 \sum_{j=1}^2 \left( C_{ij} + C_{ij,\kappa} \right).
\label{eq:costfunc}
\end{align}

\subsubsection{Optimized velocity fields}
\label{sec:sol_opt_res}

The cost function in \cref{eq:costfunc} is unfortunately not able to steer even a perfect optimisation routine towards a solution that is completely free from spurious local maxima in all cells. There are three times more contributions to the cost function than the degrees of freedom $\sigma_1\cdots\sigma_4$ in each cell---and that is when the cost function is looking for overshoots only in the Lagrange nodes, there might still be local maxima between the Lagrange nodes.

Detailed studies of the cost functions for some selected problematic cells show that there are cells where there does not exist any point in the four dimensional coefficient space $\sigma_j$ where the cost is below $\kappa_C= 1.0$. Using a brute force optimisation algorithm that explores the entire solution space would not help for those cells. In conclusion, it is likely not possible---by cell-wise slope limiting---to reconstruct a velocity field that is both solenoidal and free from local maxima based on the presented reduced basis and the chosen criteria to detect local overshoots.


\subsection{A split solenoidal slope limiting algorithm}
\label{sec:sol_lim_alg}

\begin{SCfigure}[0.8][h]
  \centering
  \includegraphics[width=0.4\textwidth]{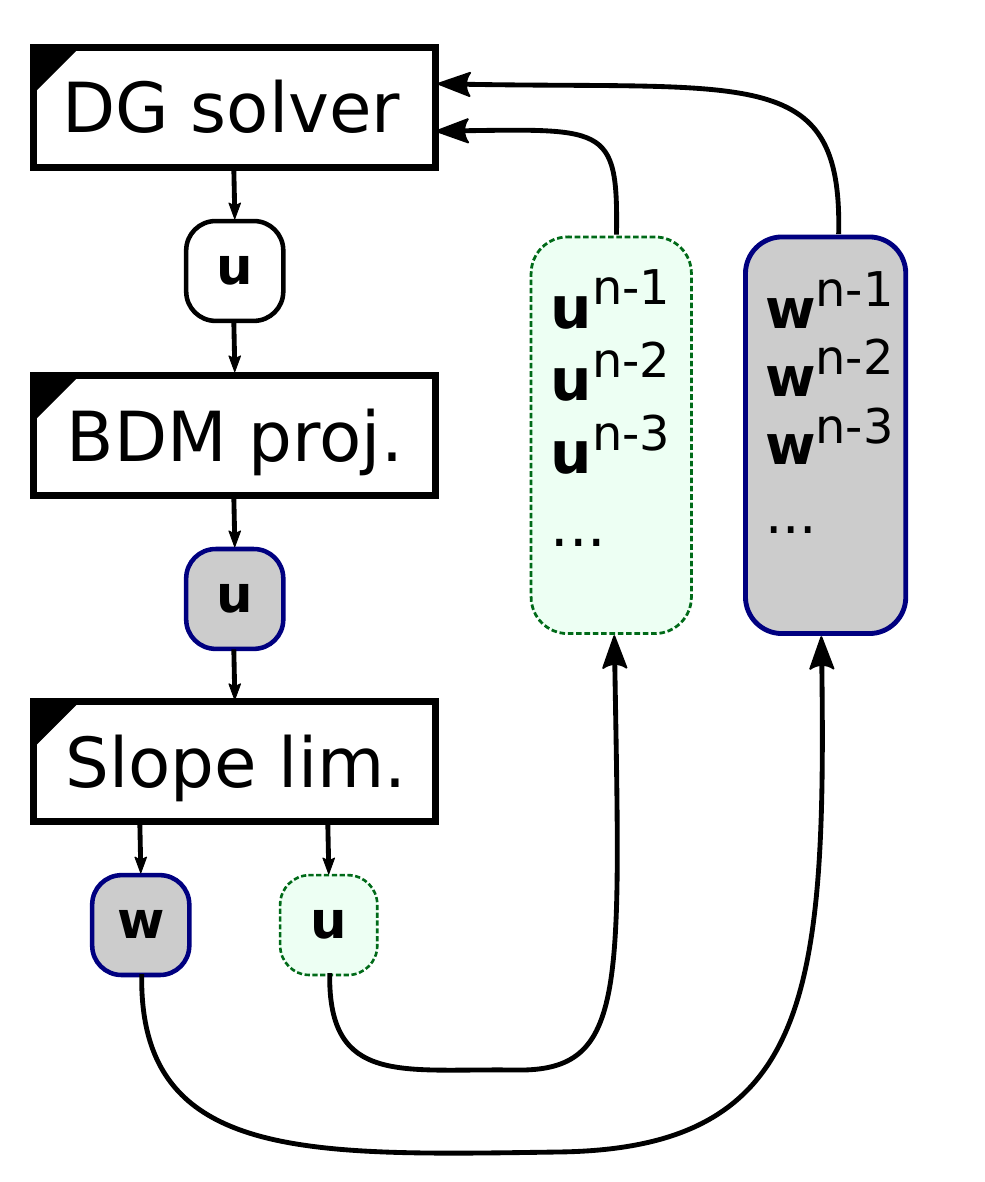}
  \caption{The slope limiting algorithm splits the velocity field into two separate time histories. Shaded velocities are solenoidal when the box boundary is a continuous line (grey background) and free from local maxima when the line is dotted (light green background).}
  \label{fig:solalg}
\end{SCfigure}

Motivated by the above findings we introduce separate limiters for the \emph{convected} velocity $\vel$ and the \emph{convecting} velocity $\wvel$. Looking at the convected and convecting velocity fields as separate, but related, is inspired by the common practice of using an explicit convecting velocity to linearise the Navier-Stokes equations---which is also what we have done in \cref{eq:ns_with_w_conv}. A similar splitting is done in cell-centered finite volume schemes where the convecting velocity field is interpolated to the facets while the unknown convected velocity field consists of cell averages. The convecting velocity can be partially slope limited in a way that does not compromise on the solenoidal properties, or left entirely unlimited. The convected velocity field must be slope limited to avoid instabilities. The overall algorithm is shown in \cref{fig:solalg}.

In the following results section we have applied three different slope limiting procedures for the velocity field. The first is the naive unlimited method, the second applies the hierarchical Taylor polynomial based limiter to each velocity component of the convected velocity, and applies no limiting to the convecting velocity. The third method applies slightly different versions of the solenoidal slope limiter from \cref{sec:sol_opt_cost} to the two velocity fields. The convecting velocity field is limited without regards for the resulting cost function value in each cell; hence, for some cells the suppression of local maxima will not be successful. The convected velocity field is treated similarly, but here the cell velocity fields are replaced with the component-wise limited fields from the hierarchical Taylor polynomial based limiter, $\vel_\text{HT}$, in the cells where the local maxima suppression was unsuccessful.

\section{Results}
\label{sec:results}

The methods described above have been implemented in Ocellaris \citep{landet_ocellaris_2017}, a two-phase solver framework which is built on top of FEniCS
\citep{logg_automated_2012}
and implemented in a mix of Python and C++. The optimiser employed is the BFGS algorithm as implemented in SciPy \citep{jones_scipy:_2001}, but for efficiency reasons we have written the optimisation algorithm along with the cost function in C++ to decrease the running time of the solver. The hierarchical Taylor based slope limiter for scalar fields has been implemented in C++ and does not contribute significantly to the running time of the solver. The rest of the Ocellaris solver is implemented in Python and depends on the C++ code generation facilities in FEniCS to utilize the computational resources optimally. The source code, input files, and scripts to reproduce the figures shown below can be found in \cite{landet_ocellaris_2017}.

In the following text we will compare three numerical algorithms, (i) a naive DG implementation without any slope limiters, (ii) a simple velocity slope limiting implementation using the scalar Taylor based limiter on each of the convected velocity components and no limiting of the convecting velocity (referred to as `Hierarchical Taylor'), and (iii) the more involved cell-wise optimisation described above where both the convected and convecting velocity fields are optimised with slightly different criteria (referred to as `Solenoidal').



To properly test the presented limiters it is important to ensure that artificially high viscosity is not a contributing factor to the method's stability. For the air/water free surface test cases we have applied $\nu=\SI{1.0e-6}{\square\m\per\s}$ for both phases, which is artificially low for the air phase. With this choice of viscosity the non-linear convective instability impacts the results early in the simulations. This is also true for, e.g., $\nu=\SI{1.0e-4}{\square\m\per\s}$, but not for very high kinematic viscosities, when $\nu$ approaches $\num{1.0}$. Note that the dynamic viscosity $\mu$---the parameter that goes into the weak form in \cref{eq:wf_ns_coupled1}---is not the same in the two phases, it has the same factor 1000 jump as the density.

\subsection{Taylor-Green vortex}
\label{sec:results_tg}
The first test case is a study of the effect of slope limiting on the spatial convergence of the numerical scheme by considering a Taylor-Green vortex where the density is constant and the solution $\vel=[u,v], p$ is given by,
\begin{align}
u &= -\sin(\pi y)\cos(\pi x)\exp(-2\pi^2\nu t) \notag\\
v &= ~~\,\sin(\pi x) \cos(\pi y)\exp(-2\pi^2\nu t)\\
p &= -\sfrac{1}{4}\,\rho(\cos{2\pi x} + \cos{2\pi y})\exp(-4\pi^2\nu t) \notag
\end{align}
for $t\in[0,1]$ on a domain $\domain = \{(x,y) \in [0,2]\times[0,2]\}$. This test case is often solved on a periodic domain, but we use Dirichlet boundary conditions for $\vel$ to test the effect of the boundary. Initial conditions are given both at $t=0$ and $t=-\Delta t$ to be able to use second order time stepping from the start. A constant time step of $\Delta t=0.01$ is used and the kinematic viscosity is $\nu=0.005$.

\begin{figure}[htb]
  \centering
  \includegraphics[trim={0 2mm 0 2mm},clip]{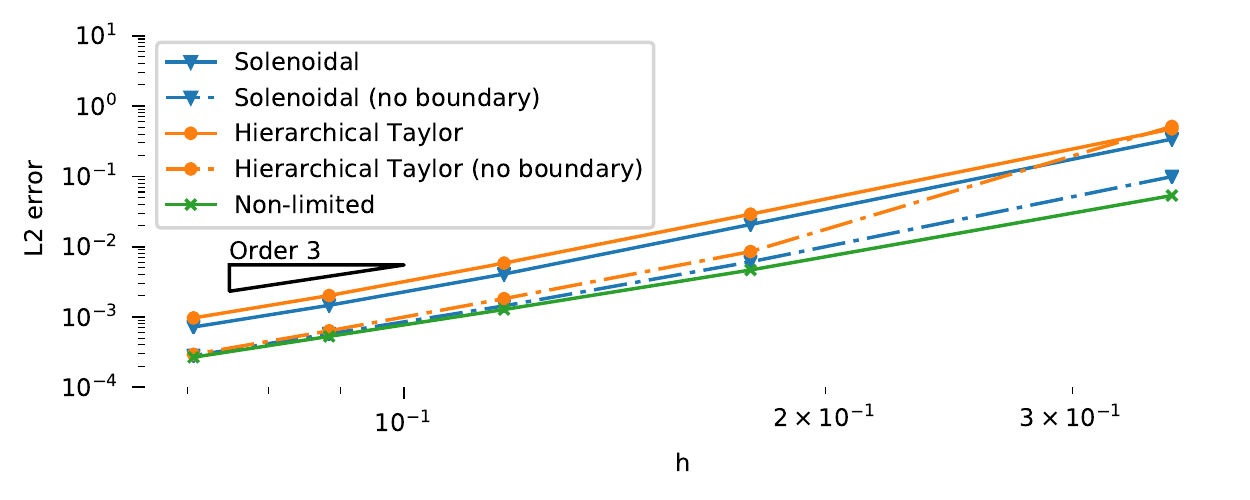}
  \caption{Spatial convergence rate on the Taylor-Green vortex test case. The rate of convergence is as expected, but missing neighbour cell info near boundaries in the limiters gives a larger constant in the $L^2$ error plot.}
  \label{fig:taylor-green}
\end{figure}

A slope limited solution should be identical to the non-limited solution for this test case---there are no non-smooth maxima in the true solution. Still, exact equivalence is not obtained since the DG FEM vector field is only a numerical approximation. The main discrepancy is in the handling of boundary conditions. Vertices on the boundary will be missing neighbour cells and may hence be unnecessarily slope limited. \Cref{fig:taylor-green} shows that the results converge towards the non-limited solution when slope limiting is avoided in the boundary facing cells. When boundary cells are included in the limiter the expected third order $L^2$ convergence rate is still obtained, but the constant is larger.

\subsection{Dam break}
\label{sec:results_dam}

\begin{figure}[h]
  \centering
  \includegraphics[trim={0 0mm 0 0mm},clip,scale=0.6]{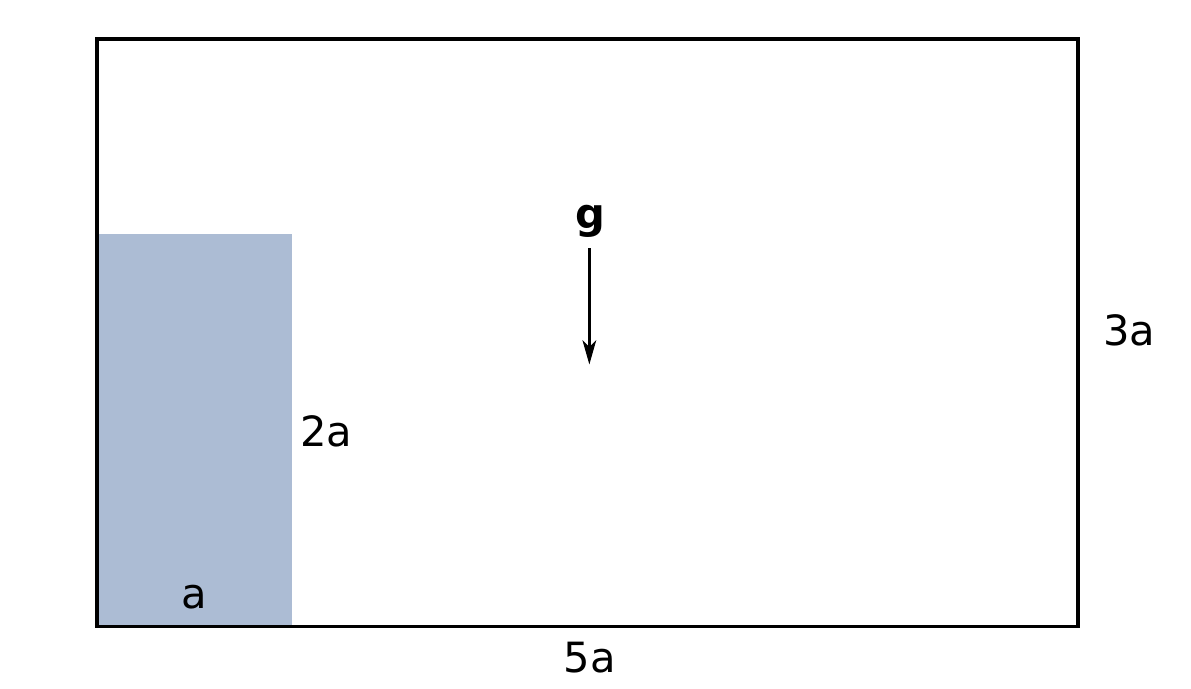}
  \caption{Dam break simulation geometry}
  \label{fig:dambreak_geom}
\end{figure}

The first two-phase flow example is a classic a dam break in a box simulation as illustrated in \cref{fig:dambreak_geom} based on the experiments by \citet{martin_part_1952}. This is one of the most commonly used test cases in the high Reynolds number, low surface tension regime. This is the regime most interesting for studying marine and offshore structures, which is the ultimate goal of the method.

The 2D rectangular water column is $1a$ wide and $2a$ high and starts at rest in the lower left corner of a box filled with air which is $5a$ wide and $3a$ high. The size of the column is the same as in the experiments, $a=\num{2.25}\,\text{in}=\SI{0.05715}{\m}$. The two phases are water and air with densities \SI{1000}{\kg\per\cubic\m} and \SI{1.0}{\kg\per\cubic\m} respectively. This corresponds to tables 2 and 6 in \citet{martin_part_1952}. The acceleration of gravity is $g=\SI{9.81}{\m\per\square\s}$ in the negative y-direction and the kinematic viscosity is $\nu=\SI{1.0e-6}{\square\m\per\s}$ for both phases. 

\begin{figure}[h]
  \centering
  \includegraphics[trim={0 2mm 0 2mm},clip]{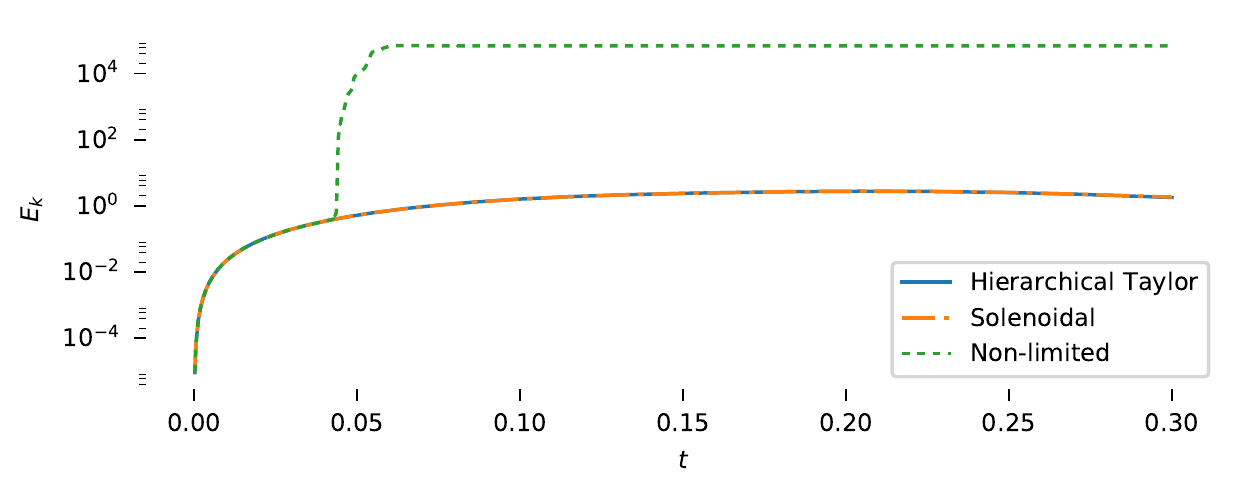}
  \caption{Kinetic energy as a function of time for three slope limiting strategies}
  \label{fig:kinen}
\end{figure}

The total kinematic energy, $E_k = \int_\domain \sfrac{1}{2}\, \rho\,\vel\cdot\vel\dif\pos$, is shown in \cref{fig:kinen} as a function of time for the slope limited and the non-limited methods. From the start, with both fluids at rest, the kinetic energy increases as the water mass starts to flow down and towards the right wall. There is a slight reduction as the water hits this wall at approximately $t=\SI{0.19}{\s}$. The Gibbs oscillations start to dominate the non-limited solution after a short time and from $t=\SI{0.04}{\s}$ the solution is non-physical and completely dominated by numerical errors. In the rest of this paper we will remove the non-limited method from the results as it leads to non-physical solutions in all cases.

\begin{figure}
  \centering
  \subfloat[Surge front position]{%
    \label{fig:compare_mm_surge}%
    \includegraphics[width=0.45\textwidth,trim={0 2mm 0 2mm},clip]{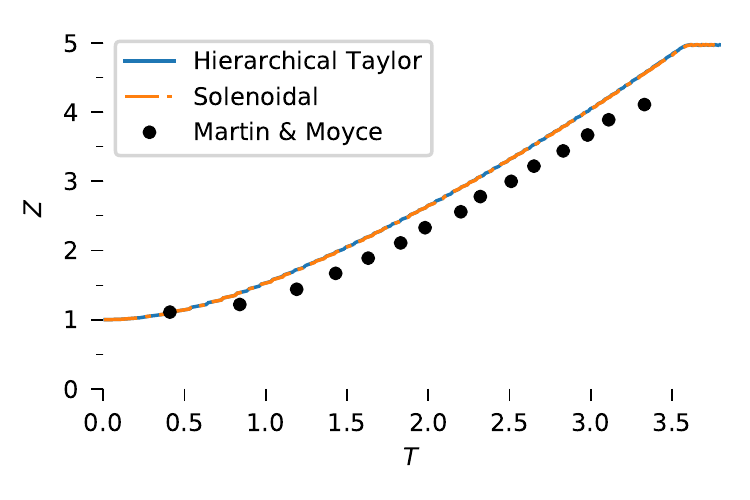}}
  ~
  \subfloat[Height of water column]{%
    \label{fig:compare_mm_height}%
    \includegraphics[width=0.45\textwidth,trim={0 2mm 0 2mm},clip]{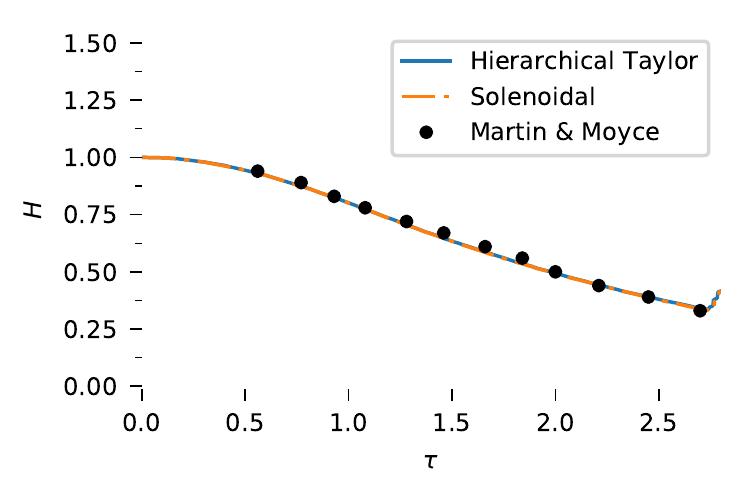}}
  
  \caption{Comparison of the numerical results with experimental results by \citet{martin_part_1952}. The vertical axis scaling is $Z=z/a$ and $H=\eta / 2a$. Measured from the lower left corner, $z$ is the surge front position and $\eta$ is the height of the water column. The horizontal axis scaling is $T=t \sqrt{2 g / a}$ and $\tau=t \sqrt{g / a}$.}
  \label{fig:compare_mm}
\end{figure}

The results of the two slope limited methods are compared to the experimental results in \cref{fig:compare_mm}. The experimental data points are the ensemble averages of the results reported for the same geometry by \citet{martin_part_1952}. We have employed free slip boundary conditions, and hence the surge front moves slightly faster than in the experimental results, but the qualitative behaviour is correct. The maximum height of the water column matches very well until the water hits the domain boundary and creates a jet shooting up along the right wall at $t\approx\SI{0.19}{\s}$ which corresponds to $T\approx3.5$ and $\tau\approx2.8$. The right wall was placed much further away in the experimental setup.

\begin{figure}
  \centering
  \includegraphics[trim={0 3mm 0 2mm},clip]{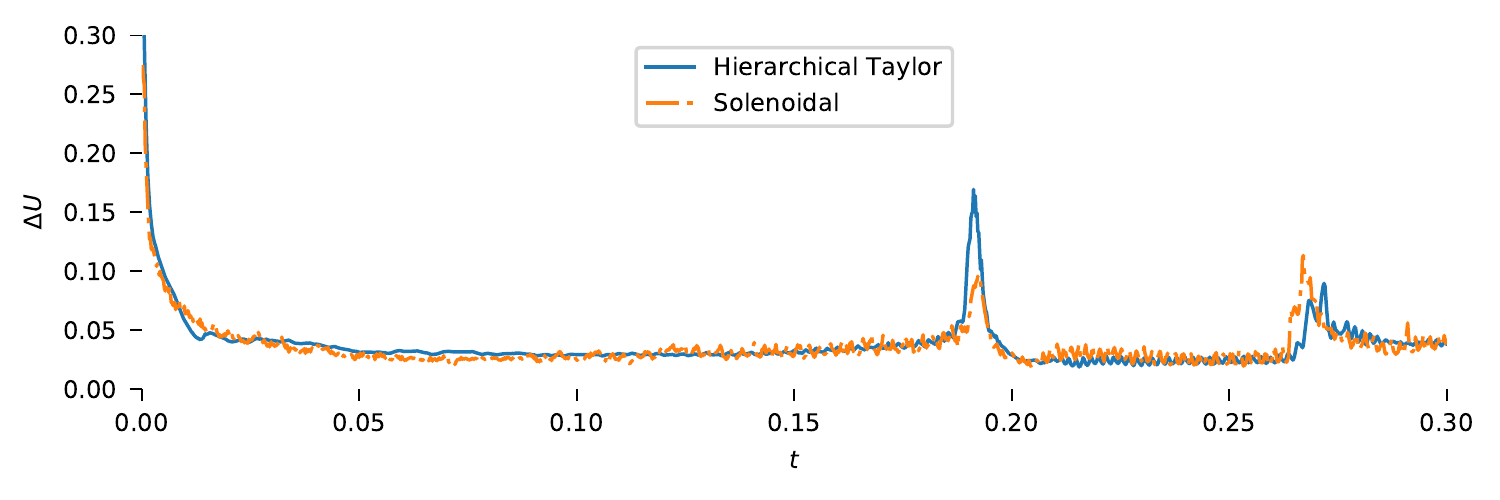}
  \caption{Dam break; comparison of the two velocity slope limiters in terms of, $\Delta U$, the scaled $L^2$ distance between convected and convecting velocity fields.}
  \label{fig:dambreak_uconv_diff}
\end{figure}

As can be seen in \cref{fig:kinen,fig:compare_mm}, the two investigated limiters perform very similarly on the dam break test case. The `Solenoidal' method requires significantly more computation per time step than the component-wise Taylor based limiter and is more complex to implement. One reason why one could still consider using this method is that the optimisation may keep the difference between the convected and the convecting velocity fields smaller and hence closer to the true solution. To test how much this optimisation  contributes we have calculated the difference between the velocity fields, $\Delta U=\norm{\vel-\wvel} / \norm{\vel}$, for all time steps and the results can be seen in \cref{fig:dambreak_uconv_diff}. 

The difference is large in the beginning since both $\vel$ and $\wvel$ are close to zero at this time. After the initial phase the difference is low for both limiters. We can see that right before the water impacts the wall at approximately $t=\SI{0.19}{\s}$ the slope limiters start to perform differently for some time until the impact is over. The reason is that the `Hierarchical Taylor' slope limiter introduces a very high divergence in the convected velocity field in the tank corner. The optimisation based limiter performs as expected and the difference between the two fields is lower, but not entirely removed.

Later in the time series there is a new event when the water is shooting up in a thin jet along the right wall at approximately $t=\SI{0.27}{\s}$. Here we can see that the optimised slope limiter is introducing slightly higher $\Delta U$ differences than the Taylor based slope limiter, quite contrary to the intention. We have no direct explanation of this, except that it seems related to the locally high shear in the vertical velocity field.

\subsection{Tank filling}
\label{sec:results_filling}

To provide a more challenging test of the method's stability with a large and complex free surface we have performed a set of tank filling simulations inspired by \citet{guermond_conservative_2017}, but unlike in their work we have used physical properties closer to normal water and air, and the results are hence more energetic. The physical properties are the same as in the dam break test case.

\begin{figure}
  \centering
  \includegraphics[width=0.8\textwidth]{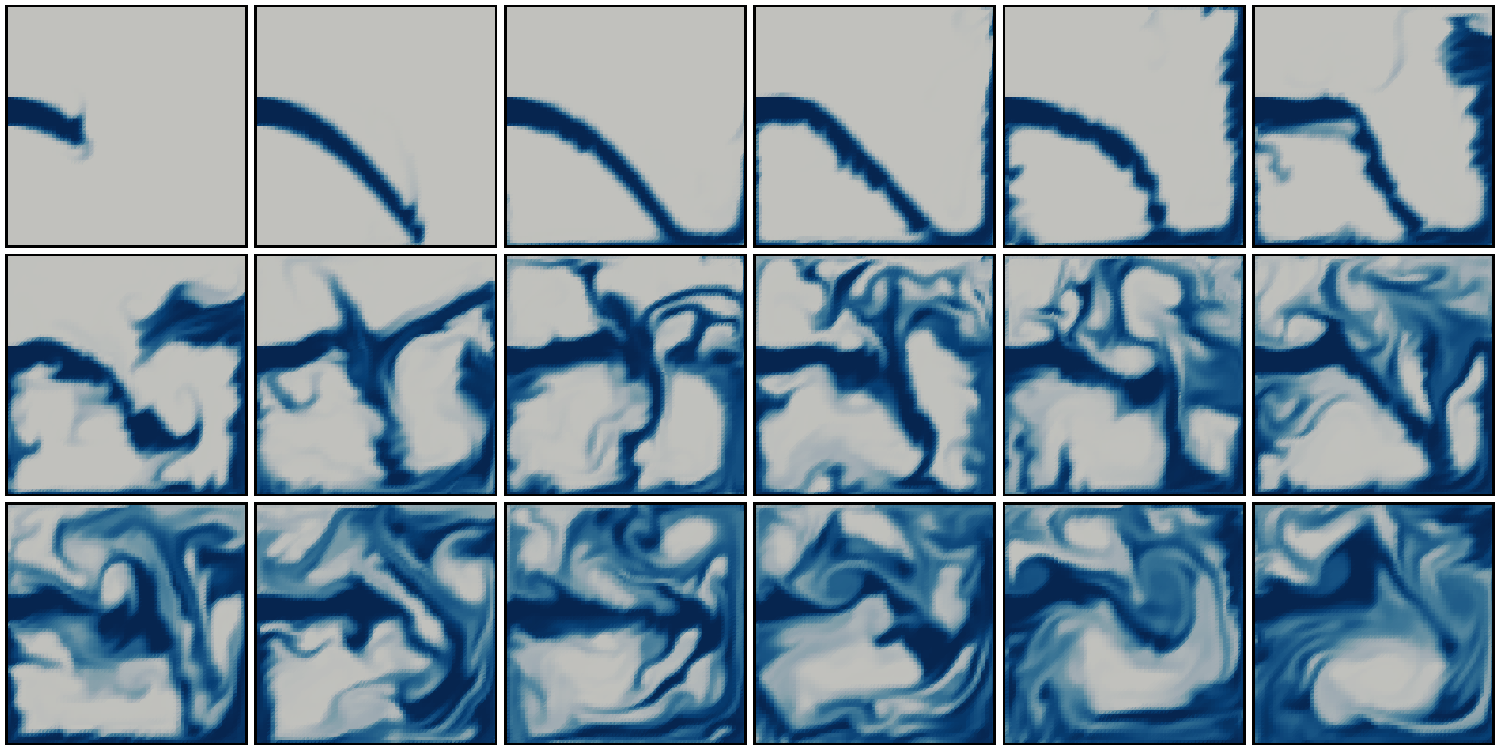}
  \caption{Tank filling results. The first row shows the density field at $t=$ 0.167, 0.333, 0.500, 0.667, 0.833, and 1.000, the second and third row have the same temporal spacing, hence the lower rightmost image shows the simulation end point, $t=3$ s. The finest resolution is shown, using the `Hierarchical Taylor' slope limiter.}
  \label{fig:tank_filling_colours}
\end{figure}

The geometry of the tank can be seen along with the resulting density field in \cref{fig:tank_filling_colours}. The domain is the unit square and the inlet is located between \SI{0.5}{\m} and \SI{0.625}{\m} above the floor on the left wall, $\SI{1/8}{\m}$ wide. The outlet is located in the roof between \SI{0.5}{\m} and \SI{0.625}{\m} from the left wall and is also $\SI{1/8}{\m}$ wide. The inlet and outlet velocities are prescribed as \SI{2}{\m\per\s}, constant for all time. At t=0 the tank is completely filled with air, the velocity inside is zero, and the pressure is hydrostatic.

The results are computed on a regular mesh with triangular cells and no mesh grading. There are eight cells across the inlet and 64x64x2 cells in total. The domain is first divided into squares and then these are subdivided into triangles. The time step is $\Delta t=\SI{0.0001}{\s}$. The solution has also been calculated on a finer mesh with 16 elements across the inlet, 128x128x2 cells in total. The time step was then $\Delta t=\SI{5.0e-05}{\s}$ and only the `Hierarchical Taylor' slope limiter was tested. The results from all three simulations qualitatively exhibit the same physics and look similar. The density field from the simulations on the fine mesh is shown in \cref{fig:tank_filling_colours} at regular intervals up to the maximum simulation time of \SI{3}{\s}.

\begin{figure}[htb]
  \centering
  \subfloat[Total mass]{%
    \label{fig:tank_filling_mass}%
    \includegraphics[width=0.45\textwidth,trim={0 2mm 0 2mm},clip]{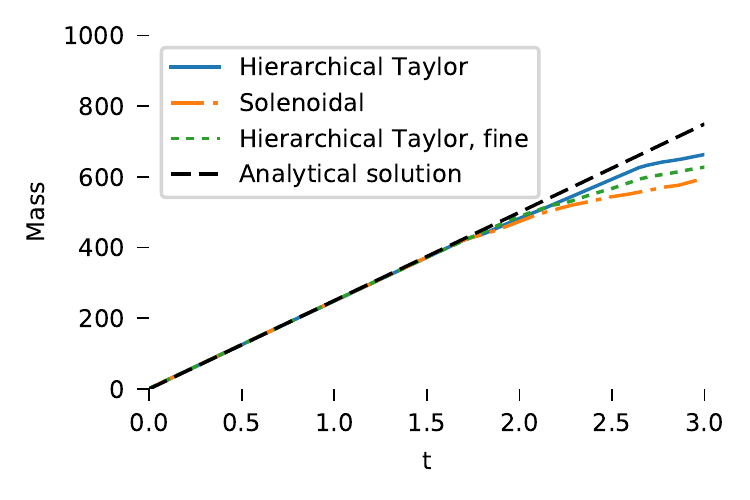}}
  ~
  \subfloat[Total energy]{%
    \label{fig:tank_filling_energy}%
    \includegraphics[width=0.45\textwidth,trim={0 2mm 0 2mm},clip]{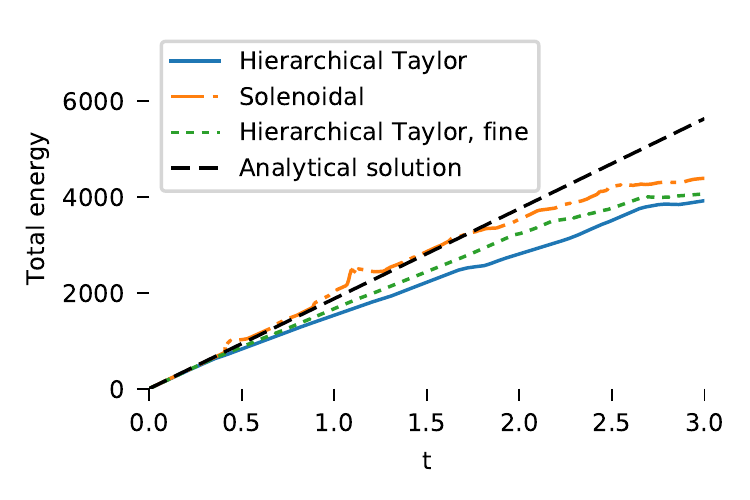}}

  \caption{Tank filling: comparison of the two velocity slope limiters in terms of conservation of mass and energy}
\end{figure}

The performance of the slope limiters is compared in terms of stability and conservation. We have calculated the total mass and energy inside the tank and the results can be seen in \cref{fig:tank_filling_mass,fig:tank_filling_energy}. The analytical solutions in the figures are computed under the assumption that only air exits through the outlet, which is not true in the second half of the simulation. When looking at conservation of mass in the first part of the simulations, only very minor differences can be seen between the methods; both methods preserve mass very well. This is exactly as expected since mass conservation is the key invariant in both slope limiters---the convecting velocity field is always solenoidal.

The loss of mass and energy through the outlet---starting at approximately 1.7 seconds---can also be seen in the time history of the total energy in \cref{fig:tank_filling_energy}. Up to that point the optimisation based solenoidal slope limiter can be seen to preserve total energy better than the component-wise hierarchical Taylor based limiter. Refining the mesh improves the results from the hierarchical Taylor based limiter, but the optimised solenoidal limiter still outperforms it, even if it is running on a much coarser mesh.

Both methods are stable for the duration of the simulation, but some spurious increases in total energy can be seen in the `Solenoidal' method. The `Hierarchical Taylor' method has no noticeable spurious increases, at the expense of slowly loosing energy. The reason for the spurious increase in energy in the `Solenoidal' method around $\SI{1.2}{\s}$ in \cref{fig:tank_filling_energy} may be the fact that local maxima are allowed to occur in between the Lagrange polynomial nodes.

\begin{figure}
  \centering
  \includegraphics[trim={0 3mm 0 2mm},clip]{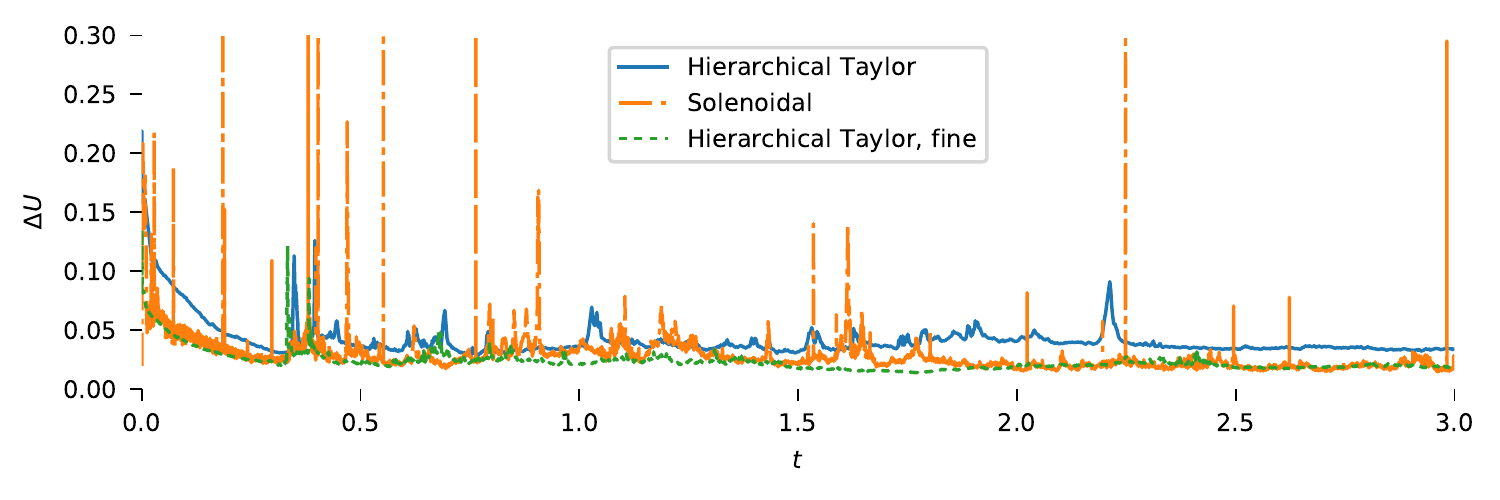}
  \caption{Tank filling: comparison of the two velocity slope limiters in terms of, $\Delta U$, the scaled $L^2$ distance between convected and convecting velocity fields.}
  \label{fig:tank_filling_uconv_diff}
\end{figure}

The difference between the convected and the convecting velocity, $\Delta U$, has been calculated in the same manner as in the dam break test case and the results can be seen in \cref{fig:tank_filling_uconv_diff}. Both methods preserve a low difference between the velocity fields for most of the time steps, but both have instances when the difference grows larger. This is particularly noticeable in the `Solenoidal' method which has some large spikes. In general the optimised method maintains a lower difference, on par with the refined hierarchical Taylor based method. Both methods are shown to recover from temporary spikes and return to a low difference, which is the correct solution. 


\section{Discussion}
\label{sec:discussion}

We have shown how convective instabilities in higher order two-phase flow simulations can be stabilised without sacrificing mass conservation. Our efforts have been focused on the use of slope limiters to achieve this stabilisation.
Using optimisation on a solenoidal reduced basis to create one single velocity field that is both stable and solenoidal was investigated. The results show strong indications that this may in fact not be possible.
To overcome this problem, two methods that treats the convected velocity field differently from the convecting velocity field were proposed. These methods combine exact mass and momentum conservation with convective stability for two-phase simulations containing large density jumps inside the computational domain.

The suggested methods have been tested to see if there were any negative impact on  the solution of smooth problems; and optimal convergence on a Taylor-Green vortex problem was show. Further, a two-dimensional dam break with a factor 1000 sharp density jump in the computational domain was tested. Such problems are not possible to handle without some form of higher order convective stabilisation. Both slope limiting strategies handled this problem well. Finally, we have studied mass and energy conservation in a very energetic flow---a tank filling simulation. A summary of our results is that slope limiting is a possible way to stabilise a mass conserving discontinuous Galerkin method for the incompressible Navier-Stokes equation with large density jumps. 


There are other methods than slope limiting for stabilising high order convective instabilities in discontinuous Galerkin methods. Instead of using an explicit post processing operator one can add specifically tailored implicit damping to the cells near the free surface with a method such as the entropy viscosity method, \citep{zingan_implementation_2013}. Another method that has been used is to selectively reduce the polynomial approximation order in cells near the free surface, a special use of $p$-adaptivity (Robert J. Labeur, private communication, 2017).
The advantages of the explicit slope limiting method is the decoupling of the convective stabilisation from the definition of the weak form and the assembly of the equation system. The relative ease of implementation is also a large advantage of the component-by-component velocity limiting method. Some type of scalar slope limiter is likely to be already present in a DG FEM code. Both the presented methods require no extra method dependent viscosity parameters to be derived, and no selective $p$-adaptivity is needed in the software. There is no need to tune parameters in the methods to achieve stability. Extension to higher order is also straight forward.

\section*{Acknowledgements}

The authors are thankful to Miroslav Kuchta for proofreading and valuable discussion related to this work.

\bibliographystyle{gCFD}
\bibliography{references}
\end{document}